%% file: main.tex
\documentclass[a4paper]{llncs}

\usepackage[usenames]{color}
\usepackage{times}
\usepackage{hyperref}
 \usepackage{latexsym}
\usepackage{amsmath}
\usepackage[linesnumbered,vlined,ruled]{algorithm2e}
\usepackage{amssymb}
\usepackage{tikz}
\usepackage{paralist}
\usepackage{subfigure}
\usepackage{listings}
\usepackage{multirow}
\usepackage{mathtools}
\usepackage{wrapfig}
\usepackage{stackrel}

\input{macros}

\title{Lazy Automata Techniques for WS1S\vspace*{-3mm}}

\author{
  Tom\'{a}\v{s} Fiedor\inst{1} \and
  Luk\'{a}\v{s} Hol\'{i}k\inst{1} \and
  Petr Jank\r{u}\inst{1} \and
  Ond\v{r}ej Leng\'{a}l\inst{1,2} \and
  Tom\'{a}\v{s} Vojnar\inst{1}
}

\institute{
  {FIT, Brno University of Technology, IT4Innovations Centre of Excellence, Czech~Republic} \and
  {Institute of Information Science, Academia Sinica, Taiwan}
}

\begin{document} 

\maketitle

\vspace*{-5mm}
\begin{abstract}
We present a new decision procedure for the logic WS1S.
It originates from the classical approach, which first builds an automaton
accepting all models of a formula and then tests whether its language is empty.
The main novelty is to test the emptiness on the fly, while constructing a
symbolic, term-based representation of the
automaton, and prune the constructed state space from parts irrelevant to the
test. 
The pruning is done by a~generalization of two techniques used in
antichain-based language inclusion and universality checking of finite automata:
subsumption and early termination.
The~richer structure of the WS1S decision problem allows us, however, to
elaborate on these techniques in novel ways. 
Our experiments show that the proposed approach can in many cases significantly
outperform the classical decision procedure (implemented in the \mona~tool) as
well as recently proposed alternatives. 
\end{abstract}

\vspace{-7.0mm}
\section{Introduction}\label{Section:Introduction}
\vspace{-1.0mm}

Weak monadic second-order logic of one successor (WS1S) 
is a~powerful language for
reasoning about regular properties of finite words.
It has found numerous uses, from software and
hardware verification through controller synthesis to computational
linguistics, and further on.
Some more recent applications of WS1S include
verification of pointer programs and deciding related logics
\cite{strand1,strand2,adam,hip/sleek,jahob} as well as synthesis from regular
specifications~\cite{regsy}.
Most of the successful applications were due to the tool \mona{}~\cite{monapaper},
which implements classical automata-based decision procedures for WS1S and WS2S
(a generalization of WS1S to finite binary trees).
The worst case complexity of WS1S is
%
nonelementary~\cite{meyer-lc-72} 
and, despite many optimizations implemented in \mona{} and other 
tools, the complexity sometimes strikes back.
Authors of methods translating their problems to WS1S/WS2S 
are then forced to either find workarounds to circumvent the complexity blowup, such as in~\cite{strand2},
or, often restricting the input of their approach,
give up translating to WS1S/WS2S altogether~\cite{kuncak:trex}.

The classical WS1S decision procedure
builds an automaton
$A_\varphi$ accepting all models of the given formula $\varphi$ in a form of
finite words, and then tests $A_\varphi$ for language emptiness.
The bottleneck of the procedure is the size of $A_\varphi$, which can be huge
due to the fact that the derivation of $A_\varphi$ involves many nested automata
product constructions and complementation steps, preceded by determinization.
The main point of this paper is to avoid the state-space explosion involved in
the classical \emph{explicit} construction by representing automata \emph{symbolically} and testing the emptiness \emph{on the fly}, while constructing
$A_\varphi$, and by omitting the state space irrelevant to the emptiness test.
This is done using two main principles: \emph{lazy evaluation} and
\emph{subsumption-based pruning}.
%
%
%
These principles have, to some degree, already appeared in the so-called antichain-based testing
of language universality and inclusion of finite
automata~\cite{wulf:antichains}.
%
%
The richer structure of the WS1S decision problem allows us, however, to
elaborate on these principles in novel ways and utilize their power even more.  

\vspace*{-1mm}\paragraph*{Overview of our algorithm.}
Our algorithm originates in the classical WS1S decision procedure as implemented in \mona, 
in which models of formulae are encoded by finite words over a~multi-track
binary alphabet where each track corresponds to a~variable of~$\varphi$.
In order to come closer to this view of formula models as words,
we replace the input formula~$\varphi$ by a \emph{language term}
$\term_\varphi$ describing the language $L_\varphi$ of all word encodings of its models. 
\enlargethispage{5mm}

In~$\term_\varphi$, the atomic formulae of~$\varphi$ are replaced by predefined
automata accepting languages of their models.
Boolean operators ($\land$, $\lor$, and $\neg$) are turned into the
corresponding set operators ($\cup$, $\cap$, and complement) over the languages
of models.
An existential quantification~$\exists X$ becomes a~sequence of two operations.
First, a~projection $\pi_{X}$ removes information about valuations of the
quantified variable $X$ from symbols of the alphabet. 
After the projection, the~resulting language $L$ may, however, encode some but not
necessarily \emph{all} encodings of the models.
In particular, encodings with some specific numbers of trailing $\zerosymb$'s,
used as a~padding, may be missing.
$\zerosymb$ here denotes the
symbol with 0 in each track.
To obtain a~language containing \emph{all} encodings of the models, $L$
must be extended to include encodings with any number of trailing $\zerosymb$'s.
This corresponds to taking the (right) $\zerosymb^*$-quotient of~$L$,
written $L - \zerosymb^*$, which is the set of all prefixes of words of~$L$ with the remaining
suffix in~$\zerosymb^*$.
We give
an example WS1S formula $\varphi$ in~(\ref{eq:examplephi}) and its
language term $\termof\varphi$ in~(\ref{eq:exampletermphi}).
The dotted operators represent operators 
\begin{wrapfigure}[3]{r}{8.3cm}
\vspace{-12mm}
\begin{minipage}{8.2cm}
  \begin{align}
  \label{eq:examplephi}
  \varphi \equiv {} &\exists X\hspace{-1mm}: \hspace{-0.4mm}\singof{X} \land (\exists Y\hspace{-1mm}: \hspace{-0.4mm}Y \hspace{-0.5mm}= \hspace{-0.5mm}X + 1) \mezera  \\
  \label{eq:exampletermphi}
          \termof \varphi \equiv {} & \tprojof X {\left\{\autof{\singof{X}} \tcap \left(\tprojof Y {\autof{{Y = X + 1}}} \tminus {\zerosymb}^*\right)\right\}} \tminus \zerosymb^*
  \vspace{-4.0mm}
  \end{align}
\end{minipage}
\label{label}
\end{wrapfigure}
over language terms.
See Fig.~\ref{fig:automata}
for the automata $\autof{\singof{X}}$ and $\autof{Y = X + 1}$.

The main novelty of our work is that we test emptiness of $L_\varphi$ directly over~$\termof\varphi$.
The term is used as a symbolic representation of the automata 
that would be explicitly constructed in the classical procedure: 
inductively to the terms structure, starting from the leaves and combining the automata of
sub-terms by standard automata constructions that implement the term operators.
Instead of first building automata and only then testing emptiness,
we test it on the fly during the construction. This offers opportunities
to prune out large portions of the state space that turn out not to be
relevant for the test.

A sub-term $\termof\psi$ of $\termof\varphi$, corresponding to a sub-formula
$\psi$, represents final states of the automaton $\autof\psi$ accepting the language encoding models of $\psi$.
Predecessors of the final states represented by $\termof\psi$ correspond to quotients of $\termof\psi$. 
All states of~$\autof\psi$ could hence be constructed by quotienting $\termof\psi$ until fixpoint.
By working with terms, our procedure can often avoid building large parts of the automata when they are
not necessary for answering the emptiness query.
For instance, when testing emptiness of the language of a term $t_1 \cup t_2$,
we adopt the \emph{lazy approach} 
(in~this particular case the so-called \emph{short-circuit evaluation}) 
and first test emptiness of the language
of~$t_1$; if it is non-empty, we do not need to process~$t_2$.
Testing language emptiness of terms arising from quantified sub-formulae is more complicated
since they translate to $-\zerosymb^*$ quotients.
We evaluate the test on $t-\zerosymb^*$ by iterating the $-\zerosymb$ quotient from $t$. 
We either conclude with the positive result as soon as one of the iteration computes a term with a non-empty language,
or with the negative one if the fixpoint of the quotient construction is reached.
%
%
The fixpoint condition is that the so-far computed quotients \emph{subsume} the newly constructed ones,
where subsumption is a relation under-approximating inclusion of languages represented by terms.
Subsumption is also used to prune the set of computed terms so that only an
\emph{antichain} of the terms maximal wrt subsumption is kept.

Besides lazy evaluation and subsumption, our approach can benefit from multiple further optimizations.
For example, it can be \emph{combined} with the \emph{explicit WS1S decision
procedure}, which can be used to transform arbitrary sub-terms of
$\term_\varphi$ to automata.
%
%
These automata can then be rather small due to minimization, which
cannot be applied in the on-the-fly approach (the automata can, however, also explode due to determinisation and product construction, hence this technique comes with a trade-off).
We also propose a novel way of \emph{utilising BDD-based encoding} of automata transition functions in the \mona{} style for computing quotients of terms.
Finally, our method can exploit various methods of \emph{logic-based
pre-processing}, such as \emph{anti-prenexing}, which, in our experience, can often
significantly reduce the search space of fixpoint
computations.

\vspace{-2.5mm}
\paragraph*{Experiments.}


We have implemented our decision procedure in a~prototype tool called \gaston{}
and compared its performance with other publicly available WS1S solvers on
benchmarks from various sources.
In the experiments, \gaston{} managed to win over all other solvers on various
parametric families of WS1S formulae that were designed---mostly by authors of
other tools---to stress-test WS1S solvers.
Moreover, \gaston{} was able to significantly outperform \mona{} and other
solvers on a number of formulae obtained from various formal verification tasks.
This shows that our approach is applicable in practice and has a great potential
to handle more complex formulae than those so far obtained in WS1S applications.
We believe that the efficiency of our approach can be pushed much further,
making WS1S scale enough for new classes of applications.

\vspace{-2.5mm}
\paragraph*{Related work.}

As already mentioned above, \mona{}~\cite{monapaper} is the usual tool of choice
for deciding WS1S formulae.  The efficiency of \mona{} stems from many
optimizations, both higher-level (such as automata minimization, the encoding of
first-order variables used in models, or the use of BDDs to encode the
transition relation of the automaton) as well as lower-level (e.g.~optimizations
of hash tables, etc.)~\cite{monasecrets,monarestrictions}.  Apart
from~\mona{}, there are other related tools based on the explicit automata
procedure, such as \osel{}~\cite{jmosel} for a related logic M2L(Str), which
implements several optimizations (such as second-order value
numbering~\cite{osel-numbering}) that allow it to outperform \mona{} on some
benchmarks (\mona{} also provides an~M2L(Str) interface on top of the WS1S decision procedure),
or the procedure using symbolic finite automata of D'Antoni \emph{et al.} in~\cite{veanes}.

Our work was originally inspired by antichain techniques for checking universality and inclusion of finite
automata~\cite{doyen:antichain,wulf:antichains,abdulla:when}, which use
symbolic computation and subsumption to prune large state spaces
arising from subset construction.
In \cite{fiedor:tacas15}, which is a starting point for the current paper, we discussed a basic idea of generalizing these techniques to a~WS1S decision procedure.
In the current paper we have turned the idea of~\cite{fiedor:tacas15} to an algorithm efficient in practice by roughly the following steps:
(1) 
reformulating the symbolic representation of automata from nested upward and downward closed sets of automata states to more intuitive language terms,
(2) 
generalizing the procedure originally restricted to formulae in the prenex normal form to arbitrary formulae,
(3) introduction of lazy evaluation, and
(4) many other important optimizations.
%

Recently, a couple of logic-based approaches for deciding WS1S
appeared.
Ganzow and Kaiser~\cite{ganzow:new} developed a~new decision procedure for the
weak mon\-a\-dic second-order logic on inductive structures, within their tool \toss{}, which is even more
general than WS$k$S.
Their approach completely avoids automata; instead, it is based on Shelah's
composition method.
The \toss{} tool is quite promising as it outperforms \mona{} on some of the
benchmarks. It, however, lacks some features in order to perform meaningful comparison on
benchmarks used in practice.
Traytel~\cite{dreytel:coalgebras}, on the other hand, uses the classical
decision procedure, recast in the framework of coalgebras.
The work focuses on testing equivalence of a~pair of formulae, which is performed by
finding a bisimulation between derivatives of the formulae. While it is shown
that it can outperform \mona{} on some simple artificial examples, the implementation
is not optimized enough and is easily outperformed by the rest of the tools on other benchmarks.

\vspace{-2mm}
\section{Preliminaries on Languages and Automata}\label{Section:Preliminaries}
\vspace{-1mm}

A \emph{word} over a finite alphabet $\Sigma$ is a finite sequence $w = a_1\cdots
a_n$, for $n\geq 0$, of symbols from~$\Sigma$.
Its $i$-th symbol $a_i$ is denoted by $w[i]$.
For $n=0$, the word is the empty word~$\epsilon$.
A~language $L$ is a~set of words over $\Sigma$.
We use the standard language operators of concatenation $L.L'$ and iteration $L^*$.
The (right) quotient of a~language~$L$ wrt~the
language~$L'$ is the language $L - L'  = \{u\mid \exists v \in L': uv\in L\}$.
We abuse notation and write $L - w$ to denote $L - \{w\}$, for a~word $w \in \Sigma^*$.

A~\emph{finite automaton} (FA) over an alphabet $\Sigma$ is a quadruple $\aut = (Q,\delta,I,F)$
where $Q$ is a~finite set of states, $\delta \subseteq Q \times \Sigma \times Q$ is a~set of
transitions, $I \subseteq
Q$ is a~set of \emph{initial} states, and $\finst{} \subseteq Q$ is a~set of
\emph{final} states.
The \emph{$\pre$-image} of a~state $q \in Q$ over $a \in \Sigma$ is the
set of states $\preof a q = \{q'\mid (q',a,q)\in\delta\}$, and it is the set $\preof a S  = \bigcup_{q \in S} \preof a q$ for a set of states $S$. 

The language $\langof{q}$ accepted \emph{at} a state $q\in Q$ is the set of words that can be
read along a~run ending in $q$, i.e.~all
words $a_1 \cdots a_n$, for $n\geq 0$, such that $\delta$ contains transitions
$(q_0,a_1,q_1),\ldots,(q_{n-1},a_n,q_{n})$ with $q_0\in I$ and
$q_{n} = q$.
The language $\langof\aut$ of $\aut$ is then the union $\bigcup_{q\in F}\langof q$ of
languages of its final states.

\vspace{-2mm}
\section{WS1S}\label{Section:WS1S}
\vspace{-1mm}

In this section, 
we give a minimalistic introduction to the \emph{weak monadic second-order logic of one
successor} (WS1S) 
and outline its explicit decision procedure based on representing sets of models as regular languages and finite automata.
See, for instance, Comon \emph{et al.}~\cite{tata} for a more thorough introduction.

\vspace{-2mm}
\subsection{Syntax and Semantics of WS1S}
\vspace{-1mm}

WS1S 
allows quantification over second-order
\emph{variables}, which we denote by upper-case letters $X, Y, \dots$, that range
over finite subsets of~$\nat_0$.
Atomic formulae are of the form
\begin{inparaenum}[(i)]
\item  $X \subseteq Y$,
\item  $\singof{X}$,
\item  $X = \{0\}$, and
\item  $X = Y + 1$.
\end{inparaenum}
Formulae are built from the atomic ones using the logical connectives
$\wedge, \vee, \neg$, and the quantifier~$\exists \varset$ where $\varset$ is a finite set of variables (we write $\exists X$ if $\varset$ is a singleton $\{X\}$).
A \emph{model} of 
a WS1S formula $\varphi(\varset)$ with the set of free variables $\varset$
is an assignment $\rho:\varset\rightarrow\powerof{\nat_0}$ of the free variables
$\varset$ of $\varphi$ to finite subsets of $\nat_0$ for which the formula is
\emph{satisfied}, written $\rho\models\varphi $. 
Satisfaction of 
atomic formulae is defined as follows:
\begin{inparaenum}[(i)]
\item  $\rho\models X \subseteq Y$ iff $\rho(X) \subseteq \rho(Y)$,
\item  $\rho\models \singof{X}   $ iff $\rho(X)$ is a~singleton set,
\item  $\rho\models X = \{0\}    $ iff $\rho(X) = \{0\}$, and
\item  $\rho\models X = Y + 1    $ iff $\rho(X) = \{x\}, \rho(Y) = \{y\}$, and $x = y + 1$.
\end{inparaenum}
Satisfaction for formulae obtained using Boolean connectives is defined as usual.
A~formula $\varphi$ is \emph{valid}, written $\models\varphi$, iff 
all assignments of its free variables to finite subsets of~$\nat_0$ are its models,
and \emph{satisfiable} if it has a~model.
Wlog~we assume that each variable in a~formula is quantified at most once.

\vspace{-4mm}
\subsection{Models as Words}\label{sec:models-as-words}
\vspace{-1mm}
\enlargethispage{3mm}

Let $\varset$ be a finite set of variables.
A \emph{symbol} $\tau$ over $\varset$ is a mapping of all variables in~$\varset$ to the set $\{0,1\}$, e.g.~$\tau = \{X_1 \mapsto 0, X_2 \mapsto 1 \}$ for $\varset =
\{X_1,X_2\}$, which we will write as $\tau = \bintrack{X_1}{X_2}{0}{1}$ below.
The set of all symbols over $\varset$ is denoted as $\Sigma_\varset$.
We use $\zerosymb$ to denote the symbol in $\Sigma_{\varset}$ that maps all
variables to~0, i.e.~$\zerosymb = \{X \mapsto 0 \mid X \in \varset\}$.

An assignment $\rho:\varset \to \powerof{\nat_0}$ may be encoded as 
a~word $w_{\rho}$ of symbols over $\varset$ in the following way:
$w_{\rho}$ contains $1$ in the $(i+1)$-st position of the row for $X$ iff $i \in X$ in~$\rho$.
Notice that there exists an infinite number of encodings of $\rho$:
the shortest encoding is $w_{\rho}^s$ of the length $n+1$, where $n$ is the largest number appearing in any of the sets that is assigned to
a~variable of $\varset$ in $\rho$, or $-1$ when all these sets are empty.
The rest of the encodings are all those corresponding to $w_{\rho}^s$ extended with an arbitrary number of
$\zerosymb$'s appended to its end.
For example,
$\bintrack{X_1}{X_2}{0}{1}$,
$\bintrack{X_1}{X_2}{00}{10}$,
$\bintrack{X_1}{X_2}{000}{100}$,
$\bintrack{X_1}{X_2}{000\dots 0}{100\dots 0}$
are all encodings of the assignment $\rho = \left\{X_1 \mapsto \emptyset, X_2 \mapsto
\{0\}\right\}$.
We use $\langof\varphi \subseteq \Sigma_\varset^*$ to denote the language of all
encodings of a~formula $\varphi$'s models, where $\varset$ are the free variables
of $\varphi$.

For two sets $\varset$ and $\varsety$ of variables and 
any two symbols
$\tau_1, \tau_2\in\Sigma_\varset$, we write $\tau_1 \sim_\varsety \tau_2$ iff $\forall X \in \varset
\setminus \varsety: \tau_1(X) = \tau_2(X)$, i.e.~the two symbols differ (at most)
in the values of variables in~$\varsety$.
The relation $\sim_\varsety$ is generalized to words such that $w_1 \sim_\varsety w_2$
iff $|w_1| = |w_2|$ and $\forall 1 \leq i \leq |w_1| : w_1[i] \sim_\varsety w_2[i]$.
For a~language
$L\subseteq \Sigma_\varset^*$, we define $\projof \varsety L$ as the language of words $w$ that are $\sim_\varsety$-equivalent with some word $w'\in L$.  
%
%
%
Seen from the point of view of encodings of sets of assignments, 
$\cylindrof \varsety L$ encodes all assignments that may differ from those encoded by $L$ (only) in the values of variables from~$\varsety$.
If $\varsety$ is disjoint with the free variables of~$\varphi$,
then $\cylindrof{\varsety}{\langof\varphi}$ corresponds to the so-called
\emph{cylindrification} of~$\langof\varphi$,
and~if it is their subset, then $\cylindrof{\varsety}{\langof\varphi}$ corresponds
to the so-called \emph{projection}~\cite{tata}.
We~use $\cylindr_Y$ to denote $\cylindr_{\{Y\}}$ for a~variable~$Y$.

\begin{wrapfigure}[6]{r}{5.1cm}
\vspace{-11.5mm}
\hspace{-5mm}
\begin{minipage}{5.5cm}
  \begin{align}
  \label{encoding:or}
  \cyllangof\allvars{\varphi\lor\psi} &= \cyllangof\allvars\varphi\cup\cyllangof\allvars\psi\\
  \label{encoding:and}
  \cyllangof\allvars{\varphi\land\psi} &= \cyllangof\allvars\varphi\cap\cyllangof\allvars\psi\\
  \label{encoding:not}
  \cyllangof\allvars{\neg\varphi} &= \Sigma_\allvars^*\setminus\cyllangof\allvars\varphi\\
  \label{encoding:exists}
  \cyllangof\allvars{\exists \varset:\varphi} &=\cylindrof{\varset}{\cyllangof\allvars\varphi}-\zerosymb^*
  \end{align}
\end{minipage}
\end{wrapfigure}
Consider formulae over the set $\allvars$ of variables.
Let $\freeof\varphi$ be the set of free variables of $\varphi$, and let
$\cyllangof\allvars\varphi = \cylindrof{\allvars\setminus\freeof\varphi}{\langof\varphi}$ be the language $\langof{\varphi}$ cylindrified wrt those variables of $\allvars$ that are not free in $\varphi$.
Let $\varphi$ and $\psi$ be formulae 
and assume that $\cyllangof\allvars\varphi$ and $\cyllangof\allvars\psi$
are languages of encodings of their models cylindrified wrt~$\allvars$.
Languages of formulae obtained from $\varphi$ and $\psi$ using logical
connectives are defined by equations~(\ref{encoding:or}) to (\ref{encoding:exists}).
Equations (\ref{encoding:or})-(\ref{encoding:not}) above are straightforward: 
Boolean connectives translate to the corresponding set operators over the universe of encodings of assignments of variables in~$\allvars$.
Existential quantification $\exists \varset:\varphi$ translates into a~composition of two language transformations.
First, $\pi_\varset$ makes the valuations of variables of $\varset$ arbitrary, 
which intuitively corresponds to forgetting everything about values of variables in $\varset$
(notice that this is a different use of $\pi_{\varset}$ than the cylindrification
since here variables of $\varset$ \emph{are} free variables of~$\varphi$).
The second step, removing suffixes of $\zerosymb$'s from the model encodings,
is necessary since $\cylindrof{\varset}{\cyllangof\allvars\varphi}$ might be missing some encodings of models of $\exists \varset:\varphi$.
For example, suppose that $\allvars = \{X,Y\}$ and
the only model of $\varphi$ is $\{X\mapsto \{0\},Y\mapsto \{1\}\}$, yielding $\cyllangof\allvars\varphi = \bintrack{X}{Y}{10}{01}\hspace{-0.6mm}
\bintracknolabbr{0}{0}^*$.
Then $\cylindrof{Y}{\cyllangof\allvars\varphi} = \bintrack{X}{Y}{10}{??}\hspace{-0.6mm}
\bintracknolabbr{0}{?}^*$
does not contain
the shortest encoding $\bintrack{X}{Y}{1}{?}$ (where each `?' denotes an
arbitrary value) of the only model $\{X\mapsto \{0\}\}$ of $\exists Y:\varphi$.
It only contains
its variants with at least one $\zerosymb$ appended to it. This
generally happens for models of $\varphi$ where the largest number in the value
of the variable~$Y$ being eliminated is larger than maximum number found in the values of
the free variables of $\exists Y:\varphi$. 
The role of the $-\zerosymb^*$ quotient is to include the missing encodings of models
with a~smaller number of trailing $\zerosymb$'s into the language.

The standard approach to decide satisfiability of a WS1S formula $\varphi$ with
the set of variables~$\allvars$ is to construct an automaton $\aut_\varphi$
accepting $\cyllangof{\allvars}\varphi$ and check emptiness of its language.
The construction starts with simple pre-defined automata $\autof\psi$ for $\varphi$'s atomic
formulae $\psi$ (see Fig.~\ref{fig:automata} for examples of automata for selected atomic formulae and
e.g.~\cite{tata} for more details) accepting cylindrified languages~$\cyllangof\allvars{\psi}$
of models of~$\psi$.
These are simple regular languages.
The construction then continues by inductively constructing automata
$\autof{\varphi'}$ accepting languages $\cyllangof\allvars{\varphi'}$ of models for all
other sub-formulae $\varphi'$ of $\varphi$, using equations (\ref{encoding:or})--(\ref{encoding:exists}) above.
The~language operators used in the rules are implemented using standard
automata-theoretic constructions (see~\cite{tata}). 


\vspace{-3.0mm}
\section{Satisfiability via Language Term Evaluation}\label{sec:term-eval}
\vspace{-2.0mm}
\enlargethispage{6mm}

This section introduces the basic version of our symbolic algorithm 
 for deciding
satisfiability of a WS1S formula $\varphi$ with a~set of variables~$\allvars$. 
Its optimized version is the subject of the next section.
To simplify presentation, we consider the particular case of \emph{ground}
formulae (i.e.~formulae without free variables), for which satisfiability
corresponds to validity. Satisfiability of a~formula with free variables can be
reduced to this case by prefixing it with existential quantification over the
free variables. If~$\varphi$ is ground, the language $\cyllangof\allvars\varphi$ is
either $\Sigma_\allvars^*$ in the case $\varphi$ is valid, or empty if $\varphi$ is
invalid. Then, to decide the validity of $\varphi$, it suffices to test if
$\epsilon\in\cyllangof\allvars\varphi$. 

Our algorithm evaluates the
so-called \emph{language term} $\termof\varphi$,
a symbolic representation of the language $\cyllangof\allvars\varphi$, whose
structure reflects the construction of $\autof\varphi$. 
It is a~(finite) term generated by the following grammar:
\vspace{-2.5mm}
\begin{equation*}
t ::=
\aut \mid
t \tcup t \mid
t \tcap t \mid
\tcmpl{t} \mid
\tprojof \varset{t} \mid
t \tminus \something \mid
t \tminus \something^* \mid
T
\vspace{-2.5mm}
\end{equation*}
where $\aut$ is a~finite automaton over the alphabet $\Sigma_\allvars$, 
$\something$ is a symbol $\tau \in \Sigma_\allvars$ or
a~set $S\subseteq\Sigma_\allvars$ of symbols,
and $T$ is a finite set of terms.
We use marked variants of the operators to distinguish the syntax of language terms manipulated by our algorithm from the cases when we wish to denote the semantical meaning of the operators.
A~term of the form $\term\tminus \something^*$ is called a~\emph{star quotient}, or shortly a~\emph{star}, and a~term~$\term\tminus \tau$ is
a~\emph{symbol quotient}. 
Both are also called \emph{quotients}.
The \emph{language $\langof\term$ of a term $\term$} is obtained by taking the
languages of the automata in its leaves and combining them using the term operators.
Terms with the same language are \emph{language-equivalent}. 
The special terms~$T$, having the form of a set, represent intermediate states
of fixpoint computations used to eliminate star quotients.%
The language of a set~$T$ equals the \emph{union} of the languages of its elements.  
The reason for having two ways of expressing a union of terms is a different treatment of $\tcup$ and $T$, which will be discussed later. 
We use the standard notion of isomorphism of two terms, extended with having two set terms isomorphic iff they contain isomorphic elements.

A~formula~$\varphi$ is 
initially 
transformed into the term $\termof\varphi$ by
replacing every atomic sub-formula~$\psi$ in~$\varphi$ by the automaton $\autof\psi$ accepting
$\cyllangof\allvars\psi$, and by replacing the logical connectives with dotted term
operators according to equations (\ref{encoding:or})--(\ref{encoding:exists}) of Section~\ref{sec:models-as-words}.
The core of our algorithm is evaluation of the $\epsilon$-membership query $\epsilon\in\termof\varphi$, which will also trigger further rewriting of the term. %
%

\begin{wrapfigure}[7]{r}{5.7cm}
\vspace{-6.0mm}
\hspace{-5mm}
\begin{minipage}{6.1cm}
  \begin{align}
  \label{mem:set}
  \hspace*{-5cm}
  \memterm{T} \mezera \textiff& \mezera 	\memterm{\term}\text{ for some } t\in T\\
  \label{mem:cup}
  \memterm{\term \tcup \term'} \mezera \textiff& \mezera 	\memterm{\term} \textor \memterm{\term'}\\
  \label{mem:cap}
  \memterm{\term \tcap \term'} \mezera \textiff& \mezera 	\memterm{\term} \textand \memterm{\term'}\\
  \label{mem:cmpl}
  \memterm{\tcmpl{\term}}	   \mezera \textiff& \mezera 	\textnot {\memterm{\term}}\\
  \label{mem:proj}
  \memterm{\tprojof{\varset}{\term}}	 \mezera \textiff& \mezera 	\memterm{\term}\\
  \label{mem:aut}
  \memterm{\aut}			 \mezera \textiff& \mezera 	\initof \aut \cap \finof \aut \neq \emptyset 
  \end{align}
\end{minipage}
\end{wrapfigure}
The $\epsilon$-membership query on a~quo\-tient-free term is evaluated using equivalences (\ref{mem:set}) to (\ref{mem:aut}).
Equivalences~(\ref{mem:set}) to~(\ref{mem:proj}) reduce tests on terms to Boolean combinations of tests on their
sub-terms and allow pushing the test towards the automata at the term's leaves.
Equivalence~(\ref{mem:aut}) then reduces it to testing intersection of the initial states $\initof \aut$ and the final states $\finof\aut$ of an automaton.
%

Equivalences (\ref{mem:set}) to (\ref{mem:proj}) do not apply to quotients, which arise 
from quantified sub-formulae
(cf.~equation (\ref{encoding:exists}) in Section~\ref{sec:models-as-words}).
A~quotient is therefore (in the basic version) first rewritten into a~lan\-gu\-age-equi\-valent
quotient-free form.
This rewriting corresponds to saturating the set of final states
of an automaton in the explicit decision procedure
with all states in their $\pre^*$-image over~$\zerosymb$.
In our procedure, we use rules~(\ref{rewrite:star0}) and~(\ref{rewrite:star}). 

\begin{wrapfigure}[2]{r}{5.8cm}
\vspace{-11.5mm}
\hspace{-4mm}
\begin{minipage}{6.1cm}
  \begin{align}
  \label{rewrite:star0}
  \tprojof \varset T \tminus \zerosymb^*  \mezera{\rar}\mezera  \tprojof \varset{T \tminus \projof \varset \zerosymb ^*}
  \end{align}
\end{minipage}
\end{wrapfigure}
Rule~(\ref{rewrite:star0})
transforms the term into a~form in which a~star quotient is applied on a~plain
set of terms rather than on a projection.
A~star quotient of a set is then eliminated using a fixpoint computation that saturates the set with all quotients of its elements wrt the set of symbols $S= \projof \varset \zerosymb$.
A~single iteration is implemented using rule~(\ref{rewrite:star}).
\vspace{-0.4mm}
\begin{wrapfigure}[3]{r}{7.7cm}
\vspace{-10mm}
\hspace{-4mm}
\begin{minipage}{7.99cm}
  \begin{align}
  \label{rewrite:star}
  { T \tminus S^*   	}					 \rar &  
  \left\{
  \begin{array}{ll}
   T   										&	\text{\ if\ } T \termsetminus S \subsumed  T		\\
   (T \cup (T \termsetminus S)) \tminus S^*	&	\text{\ otherwise } 							
  \end{array}
  \right.
  \end{align}
\end{minipage}
\end{wrapfigure}

\noindent
There, 
$T \termsetminus S$ is the set $\{\term\tminus\tau\mid\term\in T\land\tau\in S\}$ of quotients of terms in~$T$ wrt symbols of~$S$. 
(Note that (\ref{rewrite:star}) uses the identity $S^* = \{\epsilon\} \cup S^*S$.)
Termination of the fixpoint computation is decided based on the subsumption relation~$\subsumed$, which is some
 relation that under-approximates language inclusion of terms. When the
condition holds, then the language of $T$ is stable wrt quotienting by $S$, i.e. $\langof T = \langof{T \tminus S^*}$.
In the basic algorithm, we use term isomorphism for $\sqsubseteq$;
later, we provide a~more precise subsumption relation with a~good trade-off between precision and cost.
Note that an iteration of rule (\ref{rewrite:star}) can be implemented efficiently by
the standard worklist algorithm, which extends $T$ only with quotients $T' \termsetminus S$ of terms
$T'$ that were added to~$T$ in the previous~iteration.

\enlargethispage{5mm}

\begin{wrapfigure}[7]{r}{5.5cm}
\vspace{-11.0mm}
\hspace{-4mm}
\begin{minipage}{5.8cm}
  \begin{align}
  \label{rewrite:cup}
  {(t \tcup t') \tminus  \tau  }				 \mezera & {\rar} \mezera 
  { (t \tminus  \tau) \tcup (t' \tminus  \tau) } \\
  \label{rewrite:cap}
  {(t \tcap t') \tminus  \tau  }				 \mezera & {\rar} \mezera 
  { (t \tminus  \tau) \tcap (t' \tminus  \tau) }\\
  \label{rewrite:cmpl}
  {\tcmpl{t} \tminus  \tau     }				 \mezera & {\rar} \mezera  { \tcmpl{t \tminus  \tau} }  \\
  \label{rewrite:proj}
  \tprojof \varset t \tminus  \tau  \mezera  & {\rar}  \mezera  \tprojof \varset {t \tminus \projof \varset \tau}\\
  \label{rewrite:aut}
  {\aut \tminus  \tau   }	 \mezera & {\rar} \mezera  { \preof{\tau}{\aut}}
  \end{align}
\end{minipage}
\end{wrapfigure}
The set $T \termsetminus S$ introduces quotient terms of the form $\term
\tminus\tau$, for $\tau \in \Sigma_\allvars$, which
also need to be eliminated to facilitate the $\epsilon$-membership test.
This is done using rewriting rules~(\ref{rewrite:cup}) to~(\ref{rewrite:aut}),
where $\preof \tau \aut$ is $\aut$ with its set of final states $F$ replaced by $\preof\tau F$.

If $\term$ is quotient-free, then
rules~(\ref{rewrite:cup})--(\ref{rewrite:proj}) applied to $\term\tminus\tau$ push the
symbol quotient down the structure of $\term$ towards the automata in the
leaves,
where it is eliminated by rule~(\ref{rewrite:aut}).
%
Otherwise, if $t$ is not quotient-free, it can be re-written using rules
(\ref{rewrite:star0})--(\ref{rewrite:aut}).
In particular, if $t$ is a~star quotient of a quotient-free term, then the
quotient-free form of $t$ can be obtained by iterating rule (\ref{rewrite:star}),
combined with rules (\ref{rewrite:cup})--(\ref{rewrite:aut})
to transform the new terms in $T$ into a quotient-free
form. 
Finally, terms with multiple quotients can be rewritten to the quotient-free form inductively
to their structure. Every inductive step rewrites some star quotient of a
quotient-free sub-term into the quotient-free form.
Note that this procedure is bound to terminate since the terms generated by
quotienting a~star have the same structure as the original term, differing only
in the states in their leaves.
As the number of the states is finite, so is the number of the terms.

\paragraph{Example 1.}

\begin{figure}[t]
\begin{center}
\vspace{-4mm}
\includegraphics[width=\textwidth,keepaspectratio]{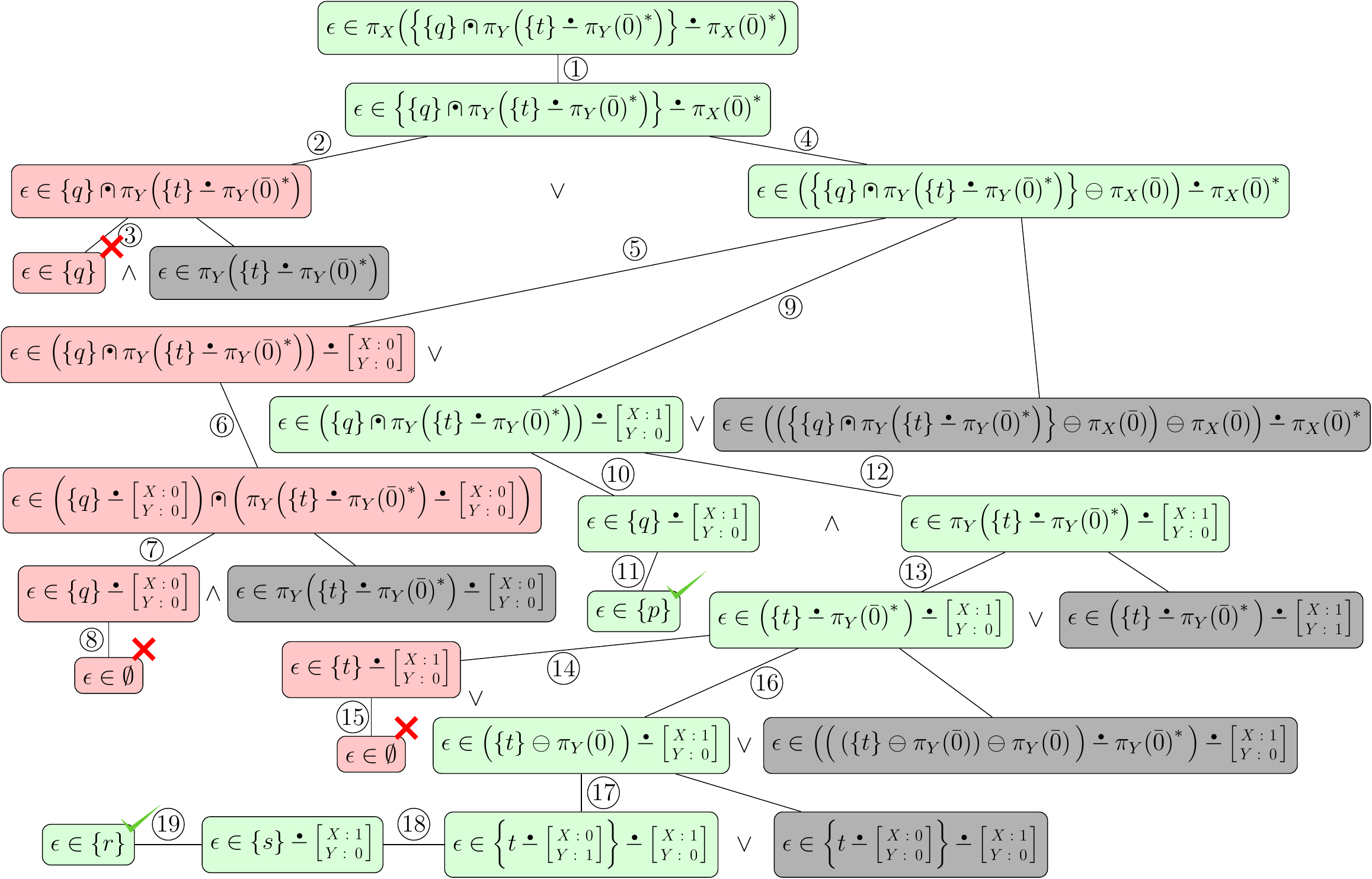}
\end{center}
\hspace{-4mm}
\vspace{-8mm}
\caption{Example of deciding validity of the formula $\varphi \equiv \exists X
: \singof{X} \land (\exists Y : Y = X + 1)$}
\label{fig:example}
\vspace{-4mm}
\end{figure}

\begin{wrapfigure}[11]{r}{3.8cm}
\vspace{-2.0mm}
\hspace{-5mm}
\begin{minipage}{4.2cm}
\begin{center}
  \includegraphics[width=3.0cm,keepaspectratio]{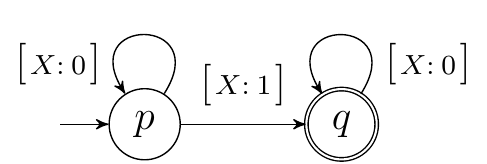}

  \vspace{-1.5mm}
  a) $\autof{\singof{X}}$
  \vspace{0.5mm}
  
  \includegraphics[width=4cm,keepaspectratio]{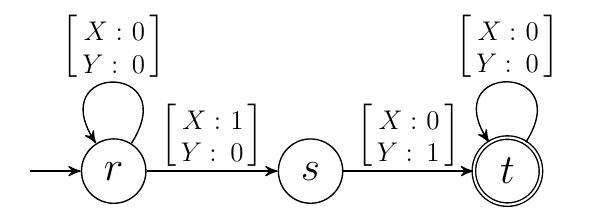}

  \vspace{-1mm}
  b) $\autof{Y = X + 1}$
\end{center}
\end{minipage}
\vspace{-4mm}
\caption{Example automata}
\label{fig:automata}
\end{wrapfigure}
We will show the workings of our procedure using
an example of testing satisfiability of the formula
  $\varphi \equiv \exists X .\,\singof{X} \land (\exists Y .\, Y = X + 1)$.
We start by rewriting~$\varphi$ into a~\emph{term}~$\termof \varphi$
representing its language~$\cyllangof \allvars \varphi$:
\vspace{-2mm}
\begin{equation*}
  \termof \varphi \equiv \tprojof X {\left\{\{q\} \tcap \tprojof Y {\{t\} \tminus \tprojof Y {\zerosymb}^*}\right\} \tminus \tprojof X {\zerosymb}^*}
\vspace{-6mm}
\end{equation*}
\enlargethispage{3mm}

\noindent
(we have already used rule~(\ref{rewrite:star0}) twice).
In the example, 
a~set~$R$ of states will denote an automaton obtained from
$\autof{\singof{X}}$ or $\autof{Y = X + 1}$  (cf.~Fig.~\ref{fig:automata})
by setting the final states to $R$.
Red nodes in the computation tree denote
$\epsilon$-membership tests that failed and green nodes those that succeeded.
Grey nodes denote tests that were not evaluated.

As noted previously, it holds that $\models \varphi$ iff $\epsilon \in \termof
\varphi$.
The sequence of computation steps for determining the $\epsilon$-membership
test is shown using the computation tree in Fig.~\ref{fig:example}.
The nodes contain $\epsilon$-membership tests on
terms and the test of each node is equivalent to a~conjunction or disjunction
of tests of its children.
Leafs of the form $\epsilon \in R$ are evaluated
as testing intersection of $R$ with the initial states of the
\mbox{corresponding automaton.}
In the example, we also use the \emph{lazy evaluation} technique (described
in Section~\ref{sec:LazyEvaluation}), which allows us to evaluate
$\epsilon$-membership tests on partially computed fixpoints.

The computation starts at the root of the tree and proceeds along the edges in the order given by their circled labels.
Edges~\stepof{2} and~\stepof{4} were obtained by a~partial unfolding of
a~fixpoint computation by rule~(\ref{rewrite:star}) and immediately applying
$\epsilon$-membership test on the obtained terms.
After step~\stepof{3}, we conclude that $\epsilon \notin \{q\}$
since $\{p\} \cap \{q\} = \emptyset$, which further refutes the whole
conjunction below~\stepof{2}, so the overall result depends on the sub-tree starting
by~\stepof{4}.
The steps~\stepof{5} and~\stepof{9} are
another application of rule~(\ref{rewrite:star}), which transforms $\projof X \zerosymb$ to
the symbols $\bintrackbr{X}{Y}{0}{0}$ and $\bintrackbr{X}{Y}{1}{0}$ respectively.
The~branch~\stepof{5} pushes the $\tminus \bintrackbr{X}{Y}{0}{0}$ quotient to
the leaf term using rules~(\ref{rewrite:cap}) and~(\ref{mem:cap}) and eventually fails because the 
predecessors of $\{q\}$ over the symbol $\bintrackbr{X}{Y}{0}{0}$ in
$\autof{\singof{X}}$ is the empty set.
On the other hand, the evaluation of the branch~\stepof{9} continues using rule~(\ref{rewrite:cap}),
succeeding in the branch~\stepof{10}.
The branch~\stepof{12} is further evaluated by 
projecting the quotient ${}\tminus \bintrackbr{X}{Y}{1}{0}$ wrt~$Y$
(rule~\ref{rewrite:proj})
and unfolding the inner star quotient
zero times (\stepof{14}, failed) and once (\stepof{16}).
The unfolding of one symbol eventually succeeds in step~\stepof{19}, 
which leads to concluding
validity of~$\varphi$.
Note that thanks to the lazy evaluation, none of the fixpoint
computations had to be fully unfolded.
\qed

\vspace{-3.0mm}
\section{An Efficient Algorithm}\label{sec:EfficientAlgorithm}
\vspace{-2.0mm}
\enlargethispage{3mm}

In this section, we show how to build an efficient algorithm based on the symbolic term rewriting approach from Section~\ref{sec:term-eval}.  
The optimization opportunities offered by the symbolic approach are to a large degree orthogonal to those of the explicit approach. 
The main difference is in the available techniques for reducing the explored automata state space.
While the explicit construction in \mona{} profits mainly from calling \emph{automata minimization} after every step of the inductive construction,
the symbolic algorithm can use generalized \emph{subsumption} and \emph{lazy evaluation}.
None of the two approaches seems to be compatible with both these techniques (at
least in their pure variant, disregarding the possibility of a combination of the two approaches discussed below).

\emph{Efficient data structures} have a major impact on performance of the
decision procedure.  The efficiency of the explicit procedure implemented in
\mona{} is to a large degree due to the BDD-based representation of automata
transition relations.  BDDs compactly represent transition functions over large
alphabets
and provide efficient implementation of operations needed in the explicit
algorithm.
Our symbolic algorithm can, on the other hand, benefit from a representation of
terms as DAGs where all occurrences of the same sub-term are represented by a
unique DAG node. 
Moreover, we assume the nodes to be associated with languages rather than
with concrete terms (allowing the term associated with a node to change during
its further processing, without a need to transform the DAG structure as long
as the language of the term does not change).

We also show that despite our algorithm 
uses a~completely different data structure than the explicit one, it can still
exploit a~BDD-based representation of transitions of the automata in the leaves
of terms.
Moreover, our symbolic algorithm can also be \emph{combined} with the explicit
algorithm.  Particularly, it turns out that, sometimes, it pays off to
translate to automata sub-formulae larger than the atomic ones.
Our procedure can then be viewed as an extension of \mona{} that takes over
once \mona{} stops managing.
Lastly, optimizations on the level of formulae often have a~huge impact on the
performance of our algorithm.
The technique that we found most helpful is the so-called \emph{anti-prenexing}.
We elaborate on all these optimizations in the rest of this section.

\vspace{-3.0mm}
\subsection{Subsumption}\label{sec:Subsumption}
\vspace{-1.0mm}
%
   Our first technique for reducing the explored
    state space is based on the notion of \emph{subsumption} between terms,
    which is 
    similar to the subsumption used in antichain-based universality and
    inclusion checking over finite automata~\cite{wulf:antichains}.
    We define subsumption as the relation~$\termsubsumed$ on terms
    that is given
    by equivalences~(\ref{subsum:set})--(\ref{subsum:aut}).
Notice that, in rule (\ref{subsum:set}), all terms of $T$ are tested against all terms of $T'$,
while in rule (\ref{subsum:cup}), the left-hand side term $t_1$ is not
tested against the right-hand side term $t_2'$ (and similarly for $t_2$ and
$t_1'$).
\begin{wrapfigure}[8]{r}{7.3cm}
\vspace{-4.0mm}
\hspace*{-3mm}
    \begin{minipage}{7.5cm}
    \begin{align}
\label{subsum:set}
      T \termsubsumed  T' &
        \textiff \forall t\in T\ \exists t'\in T':\term \termsubsumed \term' \\
\label{subsum:cup}
	  \term_1\tcup\term_2\termsubsumed\term_1'\tcup\term_2' & \textiff \term_1\termsubsumed\term_1'\textand\term_2\termsubsumed\term_2'\\
\label{subsum:cap}
	  \term_1\tcap\term_2\termsubsumed\term_1'\tcap\term_2' & \textiff \term_1\termsubsumed\term_1'\textand\term_2\termsubsumed\term_2'\\
\label{subsum:cmpl}
      \tcmpl{\term} \termsubsumed \tcmpl {\term'} &\textiff \term \termsubsumes \term'\\
\label{subsum:proj}
      \tprojof \varset {\term} \termsubsumed \tprojof \varset {\term'} & \textiff \term \termsubsumed \term'\\
\label{subsum:aut}
      \aut  \termsubsumed \aut' &\textiff  \finof\aut \subseteq \finof{\aut'}     
\end{align}
    \end{minipage}
\end{wrapfigure}
%
The reason why $\tcup$ is order-sensitive is that~the~terms~on
different sides of the $\tcup$ are assumed to be built from automata with
disjoint sets of states (originating from different sub-formulae of the original formula), and hence the subsumption test on them can never conclude positively.
    The subsumption under-approximates language inclusion and can
    therefore be used for $\subsumed$ in rule~(\ref{rewrite:star}).
It is far more precise than isomorphism and its use leads to an earlier termination of fixpoint computations. 

\begin{wrapfigure}[2]{r}{7.8cm}
\vspace{-11mm}
\hspace*{-4mm}
\begin{minipage}{8.1cm}
  \begin{align}
  \label{rewrite:prune}
  T \termequiv{ T\setminus \{t\}}   &&\text{if there is } \term'\in T\setminus \{t\} \text{ with } \term \termsubsumed \term'
  \end{align}
\end{minipage}
\end{wrapfigure}
  Moreover, $\termsubsumed$~can be used to prune star quotient terms $T \tminus S^*$ while preserving their language.
  Since the semantics of the set~$T$ is the union of the languages of its elements, 
  then elements subsumed by others can be removed while preserving the language. 
  $T$ can thus be kept in the form of an \emph{antichain} of $\termsubsumed$-incomparable terms. 
  The pruning corresponds to using the rewriting rule~(\ref{rewrite:prune}).
  %

\vspace{-2mm}
\subsection{Lazy Evaluation} \label{sec:LazyEvaluation}
\vspace{-1mm}
\enlargethispage{3mm}

The top-down nature of our technique allows us to postpone evaluation of some of
the computation branches in case the so-far evaluated part is sufficient for
determining the result of the evaluated $\epsilon$-membership or subsumption
test.
We call this optimization \emph{lazy evaluation}.
A basic variant of lazy evaluation \emph{short-circuits} elimination of quotients
from branches of $\tcup$ and $\tcap$. 
When testing whether $\epsilon \in t \tcup t'$ (rule~(\ref{mem:cup})), 
we first evaluate, e.g., the
test $\epsilon \in t$, and when it holds, we
can completely avoid exploring $t'$ and evaluating quotients there.
When testing $\epsilon \in t \tcap t'$,
we~can proceed analogously if one of the two terms is shown not to contain $\epsilon$.
Rules~(\ref{subsum:cup}) and~(\ref{subsum:cap}) offer similar opportunities for
short-circuiting evaluation of subsumption of  $\tcup$ and $\tcap$.

Let us note that subsumption is in a different position than
$\epsilon$-membership since correctness of our algorithm depends on the precision
of the $\epsilon$-membership test, but subsumption may be evaluated in any way
that under-approximates inclusion of languages of terms (and over-approximates
isomorphism in order to guarantee termination).
Hence, $\epsilon$-membership test must enforce eliminating quotients until it
can conclude the result,
while there is a~choice in the case of the~subsumption.
If subsumption is tested on quotients, it can either eliminate them,
or it can return the (safe) negative answer.
However, this choice comes with a trade-off.
Subsumption eliminating quotients is more expensive but also more precise.
The higher precision allows better pruning of the state space and earlier
termination of fixpoint computation,
which, according to our empirical experience, pays off.

Lazy evaluation can also reduce the number of iterations of a~star.
The iterations can be computed \emph{on demand}, only when required by the tests.
The idea is to try to conclude a~test $\epsilon\in T \tminus S^*$ based on the intermediate state $T$ of the fixpoint computation.
This can be done since $\langof{T}$ always under-approximates $\langof{T\tminus S^*}$, hence if $\epsilon\in\langof{T}$, then $\epsilon\in\langof{T\tminus S^*}$. 
Continuing the fixpoint computation is then unnecessary.

The above mechanism alone is, however, rather insufficient in the case of nested stars.
Assume that an inner star fixpoint computation was terminated in a state $T
\tminus S^*$
when $\epsilon$ was found in $T$ for the first time.
Every unfolding of an outer star then propagates $\tminus\tau$ quotients towards $T\tminus S^*$.
We have, however, no way of eliminating it from $(T\tminus S^*)\tminus \tau$
other than finishing the unfolding of $T\tminus S^*$ first (which eliminates the
inner star).
The need to fully unfold $T\tminus S^*$ would render the earlier lazy
evaluation of the $\epsilon$-membership test worthless.
To remove this deficiency, we need a way of eliminating the $\tminus\tau$
quotient from the intermediate state of~$T\tminus S^*$.

%
\begin{wrapfigure}[6]{r}{4.45cm}
\vspace{-11mm}
\hspace{-4mm}
\begin{minipage}{4.75cm}
  \begin{align}
  \label{rewrite:approx}
  \hspace{-1mm}T\tminus S^* &  \mezera \hspace{-0mm}  {\rar} \hspace{-0mm} \mezera 	T\tminus S^* \underapp T\\
  \label{mem:approx}
  \epsilon \in t & \underapp  t' \hspace{1mm} \mezera {\textif} \hspace{1mm} \mezera  \epsilon \in t'\\ 
  \label{subs:approx}
  t \underapp T  & \not\termsubsumed  t'  \hspace{-0.4mm} \mezera {\textif} \hspace{1mm}  \mezera  T\not\termsubsumed t'\\ 
  \label{subs:approxneg}
  t \underapp T  & \termsubsumes  t'  \hspace{-0.4mm} \mezera {\textif} \mezera \hspace{1mm}  T\termsubsumes t' 
  \end{align}
\end{minipage}
\end{wrapfigure}
The elimination is achieved by letting the star quotient $T\tminus S^*$ explicitly
``publish'' its intermediate state~$T$ using rule~(\ref{rewrite:approx}).
The symbol~$\underapp$ is read as ``\emph{is under-approximated by}.''
Rules~(\ref{mem:approx})--(\ref{subs:approxneg}) allow to conclude
$\epsilon$-membership and subsumption by testing the under-approximation on its
right-hand side (notice the distinction between ``$\textif$'' and the
``$\textiff$'' used in the rules earlier).

\begin{wrapfigure}[2]{r}{7cm}
\vspace{-12mm}
\hspace{-5mm}
\begin{minipage}{7.4cm}
  \begin{align}
  \label{rewrite:quotapprox}
  (t\underapp T) \tminus  S  \mezera  & {\rar}  \mezera   ((t\underapp T)\tminus S) \underapp T \termsetminus S
  \end{align}
\end{minipage}
\end{wrapfigure}
Symbol quotients that come from the unfolding of an outer star can be evaluated on the
approximation too using rule~(\ref{rewrite:quotapprox}), which
%
%
then applies the symbol-set quotient on the approximation~$T$ of the inner term~$t$,
and publishes the result on the right-hand side of~$\underapp$.
The left-hand side still remembers the original term~$t \tminus S$.
%

\begin{wrapfigure}[2]{r}{2.9cm}
\vspace{-11mm}
\hspace{-4mm}
\begin{minipage}{3.2cm}
  \begin{align}
  \label{rewrite:rmapprox}
  T \underapp T'  \mezera & {\rar}  \mezera 	T
  \end{align}
\end{minipage}
\end{wrapfigure}
Terms arising from rules~(\ref{rewrite:approx}) and~(\ref{rewrite:quotapprox})
allow an efficient update in the case an inner term~$\term$ spawns a~new, more
precise approximation.
In the process,
rule~(\ref{rewrite:rmapprox})
is used to remove old outdated approximations. 

\enlargethispage{4mm}

We will explain the working of the rules and their efficient implementation on
an evaluation from Example~1.
Note that in Example~1, the partial unfoldings of the fixpoints that are tested
for $\epsilon$-membership 
are
under-approxima\-tions of a~star quotient term.
For instance, branch~\stepof{14} corresponds to testing $\epsilon$-membership
in the right-most approximation of the term
$\left(\left((\{t\} \tminus \projof Y
\zerosymb ^*)\underapp \{t\}\right) \tminus 
\bintrackbr{X}{Y}{1}{0}\right) \underapp \{t\} \tminus 
\bintrackbr{X}{Y}{1}{0}$
by rule~(\ref{mem:approx})
(the branch determines
that $\epsilon \notin \{t\} \tminus \bintrackbr{X}{Y}{1}{0}$).
The result of~\stepof{14} cannot conclude the top-level $\epsilon$-membership
test because $\{t\} \tminus \bintrackbr{X}{Y}{1}{0}$ is just an
under-approximation of $(\{t\} \tminus \projof Y \zerosymb ^*)\tminus
\bintrackbr{X}{Y}{1}{0}$.
Therefore, we need to compute a~better approximation of
the term
and try to conclude the test
on it.
We compute it by first applying rule~(\ref{rewrite:rmapprox}) twice to discard
obsolete approximations ($\{t\}$ and $\{t\} \tminus \bintrackbr{X}{Y}{1}{0}$),
followed by applying rule~(\ref{rewrite:star}) to replace
$(\{t\} \tminus \projof Y \zerosymb ^*) \tminus \bintrackbr{X}{Y}{1}{0}$
with
$\left((\{t\} \cup (\{t\} \termsetminus \projof Y \zerosymb)) \tminus \projof Y
\zerosymb ^*\right) \tminus \bintrackbr{X}{Y}{1}{0}$.
Let $\beta = \{t\} \cup (\{t\} \termsetminus \projof Y \zerosymb)$.
Then, using rules~(\ref{rewrite:approx}) and~(\ref{rewrite:quotapprox}), we can
rewrite the term
$
\left(
\beta \tminus \projof Y \zerosymb ^*
\right) \tminus \bintrackbr{X}{Y}{1}{0}
$
into
$
\left(\left(
\beta \tminus \projof Y \zerosymb ^* \underapp
\beta \right) \tminus \bintrackbr{X}{Y}{1}{0}
\right)
\underapp
\beta \termsetminus \bintrackbr{X}{Y}{1}{0}
$,
where 
$\beta \termsetminus \bintrackbr{X}{Y}{1}{0}$ is the approximation used in
step~\stepof{16}, and re-evaluate the $\epsilon$-membership test on it.

Implemented na\"{i}vely, the computation of subsequent approximations of
fixpoints would involve a~lot of redundancy, e.g., in
$\beta \tminus \bintrackbr{X}{Y}{1}{0}$
we would need to recompute the term $\{t\} \tminus \bintrackbr{X}{Y}{1}{0}$,
which was already computed in step~\stepof{15}.
The mechanism can, however, be implemented efficiently so that it completely
avoids the redundant computations.
Firstly, we can maintain a~cache of already evaluated terms and never evaluate
the same term repeatedly.
Secondly,
suppose that a~term $t \tminus S^*$ has been unfolded several times into
intermediate states $(T_1 = \{t\}) \tminus S^*, T_2 \tminus S^*, \ldots, T_n \tminus S^*$.
One more unfolding using~(\ref{rewrite:star}) would rewrite $T_n \tminus S^*$ into
$T_{n+1} = (T_n \cup (T_n \termsetminus S)) \tminus S^*$.
When computing the set $T_n \termsetminus S$, however, we do not need to consider the
whole set $T_n$, but only those elements that are in $T_n$ and are not in
$T_{n-1}$ (since $T_n = T_{n-1} \cup (T_{n-1} \termsetminus S)$, all elements
of $T_{n-1} \termsetminus S$ are already in $T_n$).
%
%
%
Thirdly, in the DAG representation of terms described in
Section~\ref{sec:EfficientData}, a~term $(T \cup (T \termsetminus S)) \tminus S^*
\underapp T \cup (T \termsetminus S)$ is represented by the set of terms obtained by
evaluating $T \termsetminus S$, a~pointer to the term $T\tminus S^*$ (or rather to its
associated DAG node),
and the set of symbols $S$. The cost of keeping the history of quotienting
together with the under-approximation (on the right-hand side of $\underapp$) is hence only a pointer and a set of symbols.

\vspace{-2.0mm}
\subsection{Efficient Data Structures}\label{sec:EfficientData}
\vspace{-1.0mm}

We describe two important techniques used in our implementation that concern (1)~representation of terms and (2)~utilisation of BDD-based symbolic representation of transition functions of automata in the leaves of the terms.

\vspace{-2mm}
\paragraph{Representation of language terms.}
We keep the term in the form of a~DAG such that
all isomorphic instances of the same term are represented as a~unique DAG
node, and, moreover, when a term is rewritten
into a~language-equivalent one, it is still associated with the same DAG node.
Newly computed sub-terms are always first compared against the existing ones,
and, if possible, associated with an existing DAG node of an existing isomorphic term.
The fact that isomorphic terms are always represented by the same DAG node makes
it possible to test isomorphism of a new and previously processed term
efficiently---it is enough to test that their direct sub-terms are represented
by identical DAG nodes
(let us note that we do not look for language equivalent terms
because of the high cost of such a check).

We also cache results of membership and subsumption queries. 
The key to the cache is the identity of DAG nodes, not the represented sub-terms, 
which has the advantage that results of tests over a term are available in the
cache even after it is rewritten according to $\rar$ (as it is still represented by the same DAG node).
The cache together with the DAG representation is especially efficient when
evaluating a~new subsumption or $\epsilon$-membership test since although the result is
not in the cache, the results for its sub-terms often are.
We also maintain the cache of subsumptions closed under transitivity.

\enlargethispage{4mm}

\vspace{-2mm}
\paragraph{BDD-based symbolic automata.}
Coping with large sets of symbols is central for our algorithm. 
Notice that rules~(\ref{rewrite:star}) and~(\ref{rewrite:proj})
compute a quotient for each of the symbols in the set $\projof \varset \tau$
separately. Since the number of the symbols is $2^{|\varset|}$, this can easily make the computation infeasible.

\mona{} resolves this by using a BDD-based symbolic representation of transition relations of automata as follows:
The alphabet symbols of the automata are assignments of Boolean values to the free variables $X_1,\ldots,X_n$ 
of a formula.
The transitions leading from a state $q$ can be expressed as a function
$f_q:2^{\{X_1, \ldots, X_n\}}\rightarrow Q$
from all assignments to states such that  
$(q,\tau,q')\in\delta_q$ iff $f_q(\tau) = q'$.
The function~$f_q$ is encoded as a~multi-terminal BDD (MTBDD) with variables $X_1,\ldots,X_n$ and terminals from the set~$Q$
(essentially, it is a DAG where a path from the root to a leaf encodes a~set of transitions).
The BDD \texttt{apply} operation is then used to efficiently implement the
computation of successors of a~state via a~large set of symbols, and to
facilitate essential constructions such as product, determinization, and minimization.
We use \mona{} to create automata in leaves of our language terms.
To fully utilize their BDD-based symbolic representation,
we had to overcome the following two problems.

First, our algorithm computes predecessors of states, 
while the BDDs of \mona{} are meant to compute successors. 
To use \texttt{apply} to compute backwards,
the BDDs would have to be turned into a representation of the inverted transition function.  
This is costly and, according to our experience, prone to produce much larger BDDs. 
We have resolved this by only inverting the edges of the original BDDs and by
implementing a~variant of \texttt{apply} that runs upwards from
the leaves of the original BDDs, against the direction of the original BDD edges.
It cannot be as efficient as the normal \texttt{apply} because, unlike standard BDDs, the DAG that arises by inverting BDD edges is nondeterministic, which brings complications.  
Nevertheless, it still allows an efficient implementation of $\pre$ that works well in our implementation.

A more fundamental problem we are facing is that our algorithm can use
\texttt{apply} to compute predecessors over 
the compact representation provided by BDDs only on the level of explicit
automata in the leaves of terms.
The symbols generated by projection during evaluation of complex terms must be, on the
contrary, enumerated explicitly.
For instance, the projection $\projof \varset t$ with $\varset = \{X_1,\ldots,X_n\}$ 
 generates $2^n$ symbols, with no obvious option for reduction.
The idea to overcome this explosion is to treat nodes of BDDs as regular automata states.
Intuitively, this means replacing words over $\Sigma_{\varset}$ that encode models
of formulae by words over the alphabet 
$\{0,1\}$: 
every symbol $\tau \in
\Sigma_{\varset}$ is replaced by the \emph{string} 
$\tau$ over $\{0,1\}$. 
Then, instead of computing a quotient over, e.g., the set $\projof {\varset} \zerosymb$ of the size $2^n$, 
we compute only quotients over the $0$'s and $1$'s.
Each quotienting takes us only one level down in the BDDs representing the transition relation of the automata in the leaves of the term.
For every variable $X_i$, 
we obtain terms over nodes on the $i$-th level of the BDDs as $-0$ and $-1$ quotients of the terms at the level $i-1$.
The maximum number of terms in each level is thus $2^i$.
In the worst case, this causes roughly the same blow-up as when enumerating the ``long'' symbols. 
The advantage of this techniques is, however, that the blow-up 
can now be dramatically reduced by using subsumption to prune sets of
terms on the individual BDD levels. 

\vspace{-2.0mm}
\subsection{Combination of Symbolic and Explicit Algorithms}\label{sec:MonaSubformulae}
\vspace{-1.0mm}

It is possible to replace sub-terms of a~language term by a~language-equivalent automaton built by the explicit algorithm before starting the symbolic algorithm.  
The main benefit of this is that the explicitly constructed automata have a~simpler flat structure and can be minimized.
The minimization, however, requires to explicitly construct the whole automaton, 
which might, despite the benefit of minimization, be a~too large overhead.
The combination hence represents a~trade-off
between the lazy evaluation and subsumption of the symbolic algorithm, 
and minimization and flat automata structure of the explicit one.
The overall effect depends on the strategy of choice of the sub-formulae to be
translated into automata, and, of~course, on the efficiency of the
implementation of the explicit algorithm (where we can leverage the extremely
efficient implementation of~\mona{}).
We mention one particular strategy for choosing sub-formulae in Section~\ref{sec:Experiments}.

\enlargethispage{4mm}

\vspace{-3.0mm}
\subsection{Anti-prenexing}\label{sec:antiprenex}
\vspace{-1.0mm}
Before rewriting an input formula to a~symbolic term, we pre-process
the formula by moving quantifiers down by several language-preserving identities (which we call \emph{anti-pre\-nex\-ing}).
We, e.g., change $\exists X .\ (\varphi \land \psi)$ into $\varphi \land (\exists X.\ \psi)$ if $X$ is not free in $\varphi$.
Moving a~quantifier down in the abstract syntax tree of a~formula speeds up the fixpoint computation induced by the
quantifier.
In effect, one costlier fixpoint computation is replaced by
several cheaper computations in the sub-formulae.
This is almost always helpful since
if the original fixpoint computation unfolds, e.g., a union of two terms,
the two fixpoint computations obtained by anti-prenexing will each unfold only one operand of the union. 
The number of union terms in the original fixpoint is roughly the product of the numbers of terms in the 
simpler fixpoints.
Further,
in order to push quantifiers even deeper into the formula, we reorder the formula
by several heuristics
(e.g.~group sub-formulae with free occurrences of the same variable in a~large conjunction)
and move negations down in the structure towards
the leaves \mbox{using De Morgan's laws}.

\vspace*{-4.0mm}
\section{Experiments}\label{sec:Experiments}
\vspace*{-3.0mm}
\enlargethispage{0mm}

We have implemented the proposed approach in a~prototype tool
\gaston{}%
\footnote{The name was chosen to pay homage to Gaston, an Africa-born
brown fur seal who escaped the Prague Zoo during the floods in 2002 and made a
heroic journey for freedom of over 300\,km to Dresden. There he was caught and
subsequently died due to exhaustion and infection.},
%
%
Our tool uses the front-end of \mona{} to parse input formulae, to construct
their abstract syntax trees, and also to construct automata for sub-formulae
(as~mentioned in Section~\ref{sec:MonaSubformulae}).
%
From several heuristics for choosing the sub-formulae to be converted
to automata by \mona{},
we converged to converting only quantifier free sub-formulae and negations
of innermost quantifiers to automata since \mona{} can usually handle them
without any explosion.
%
%
\gaston{}, together with all the benchmarks described below and their detailed
results, is freely
available~\cite{gastonsite}.

%
%

\begin{wraptable}[14]{r}{40mm}
\scriptsize
\vspace*{-11.5mm}
\caption{\uabebench{} experiments}
\hspace*{-2mm}
\begin{minipage}{42cm}
  \begin{tabular}{|l||r|r||r|r|}
  \hline
  \multicolumn{1}{| c ||}{\multirow{2}{*}{\textbf{Formula}}}     &  \multicolumn{2}{c||}{\mona{}} &  \multicolumn{2}{c|}{\gaston}\\ \cline{2-5}
   & \multicolumn{1}{c|}{\textbf{Time}} & \multicolumn{1}{c||}{\textbf{Space}} & \multicolumn{1}{c|}{\textbf{Time}} & \multicolumn{1}{c|}{\textbf{Space}} \\
  \hline \hline
  \input{uabe-pclengal-dig.tex}
  \hline
  \end{tabular}
\end{minipage}
\label{tab:results-UABE}
\end{wraptable}
We compared \gaston{}'s performance with that of \mona{},
\dwina{}~implementing our older approach~\cite{fiedor:tacas15}, \toss{}
implementing the method of~\cite{ganzow:new}, and the implementations of
the decision procedures of~\cite{dreytel:coalgebras} and~\cite{veanes} (which we denote 
as \traytel{} and \veanes{}, respectively).\footnote{We are not
comparing with \osel{}~\cite{jmosel} as we did not find it
available on the Internet.}
In our experiments, we consider formulae obtained from various formal
verification tasks as well as parametric families of formulae designed to
stress-test WS1S decision procedures.\footnote{We note that \gaston{} currently
does not perform well on formulae with many Boolean variables and M2L
formulae appearing in benchmarks such as \secretsbench~\cite{monasecrets} or 
\strandbenchone~\cite{strand1,strandbench}, which are not included in
our experiments. To handle such formulae, further optimizations of \gaston{}
such as \mona{}'s treatment of Boolean variables via a dedicated
transition are needed.}
%
We performed the experiments on a~machine with the
Intel Core i7-2600@3.4\,GHz processor and 16\,GiB RAM running Debian GNU/Linux.


\enlargethispage{2mm}

\begin{wraptable}[8]{r}{60mm}
\scriptsize
\vspace*{-14.5mm}
\caption{\strandbenchtwo{} experiments}
\hspace{-2mm}
\begin{minipage}{62cm}
  \begin{tabular}{|l||r|r||r|r|}
  \hline
  \multicolumn{1}{| c ||}{\multirow{2}{*}{\textbf{Formula}}}     &  \multicolumn{2}{c||}{\mona{}} &  \multicolumn{2}{c|}{\gaston}\\ \cline{2-5}
   & \multicolumn{1}{c|}{\textbf{Time}} & \multicolumn{1}{c||}{\textbf{Space}} & \multicolumn{1}{c|}{\textbf{Time}} & \multicolumn{1}{c|}{\textbf{Space}} \\
  \hline \hline
  \input{strand-new-pclengal-dig.tex}
  \hline
  \end{tabular}
\end{minipage}
\label{tab:results-STRAND}
\end{wraptable}
Table~\ref{tab:results-UABE} contains results of our experiments with formulae
from the recent work~\cite{uabe} (denoted as \uabebench{} below), which uses
WS1S to reason about programs with unbounded arrays. 
Table~\ref{tab:results-STRAND} gives results of our experiments with
formulae derived from the WS1S-based shape analysis of~\cite{strand2} (denoted
as \strandbenchtwo{}).
%
%
In the table, we use 
\texttt{sl} to denote \strandbenchtwo{} formulae over
sorted lists and \texttt{bs} for formulae from verification of the
bubble sort procedure.
For this set of experiments, we considered \mona{} and \gaston{} only
since the other tools were missing features (e.g., atomic predicates) needed to
handle the formulae.
In the \uabebench{} benchmark, \gaston{} was used with the last optimization of
Section~\ref{sec:EfficientData} (treating MTBDD nodes as automata states) to
efficiently handle quantifiers over large numbers of variables.
In particular, without the optimization, \gaston{} hit 11 more timeouts.
On the other hand, this optimization was not efficient (and hence not used)
in \strandbenchtwo{}.


The tables compare the overall time (in seconds) the tools needed to decide the
formulae, and they also try to characterize the sizes of the generated state
spaces. 
For the latter, we count the overall number of states of the generated automata
for~\mona{}, 
and the overall number of generated sub-terms for \gaston{}.
The tables contain just a~part of the results, more can be found
in~\cite{gastonsite}.
We use \timeout{} in case the running time exceeded 2 minutes, \oom{} to denote
that the tool ran out of memory, +$k$ to denote that we added $k$ quantifier
alternations to the original benchmark, and \unsupportedbench{} to denote that
the benchmark 
requires some feature or atomic predicate unsupported by the given
tool.
On \strandbenchtwo{}, \gaston{} is mostly comparable, in two cases better, and
in one case worse than \mona{}.
On \uabebench{}, \gaston{} outperformed \mona{} on six out of twenty-three
benchmarks, it was worse on ten formulae, and comparable on the rest.
The results thus confirm that our approach can defeat \mona{} in practice.

\begin{wraptable}[8]{r}{7.70cm}
\scriptsize
\vspace*{-10mm}
\caption{Experiments with parametric families of formulae}
\hspace{-1.8mm}
\begin{minipage}{7.75cm}
  \begin{tabular}{|l|l||r|r|r|r|r|r|}
  \hline
  \multicolumn{1}{|c|}{\textbf{Benchmark}} & \multicolumn{1}{|c||}{\textbf{Src}} &
  \multicolumn{1}{|c|}{\mona} & \multicolumn{1}{|c|}{\dwina} & \multicolumn{1}{|c|}{\toss} & 
  \multicolumn{1}{|c|}{\traytel} & \multicolumn{1}{|c|}{\veanes} & 
  \multicolumn{1}{|c|}{\gaston} \\
  \hline\hline
  \input{generated-dig.tex}
  \hline
  \end{tabular}
\end{minipage}
\label{tab:results-overall}
\vspace*{-4mm}
\end{wraptable}
The second part of our experiments concerns parametric families of WS1S
formulae used for evaluation in~\cite{ganzow:new,fiedor:tacas15,veanes},
and also parameterized versions of selected \uabebench{} formulae~\cite{uabe}.
Each of these families has one parameter (whose meaning is explained in the
respective works).
Table~\ref{tab:results-overall} gives times needed to decide instances of the
formulae for the parameter having value 20.
If the tools did not manage this value of the parameter, we give in parentheses the
highest value of the parameter for which the tools succeeded.
More results are available in~\cite{gastonsite}.
In this set of experiments, \gaston{} managed to win over the other tools on
many of their own benchmark formulae.
%
In the first six rows of Table~\ref{tab:results-overall}, 
the superior efficiency of \gaston{} was caused mainly by anti-prenexing.
It turns out that this optimization of the input formula is universally effective.
When run on anti-prenexed formulae, the performance of the other tools was comparable to that of \gaston{}.
The last two benchmarks (parameterized versions of formulae from \uabebench) show, however, that \gaston{}'s performance does not stand on anti-prenexing only. 
Despite that its effect here was negligable (similarly as for all the original benchmarks from \uabebench{} and \strandbenchtwo), \gaston{} still clearly outperformed \mona{}.
We could not compare with other tools on these formulae
due to a missing support of the used features (e.g. constants). 
%

\enlargethispage{4mm}

\newpage
\vspace{-3.0mm}
\section{Concluding Remarks}\label{sec:label}
\vspace{-2.0mm}
\enlargethispage{5mm}

We have presented a novel WS1S decision procedure based on symbolic, term-based
representation of the languages of formulae. 
%
Our experiments proved that the approach is competitive and often better than
state-of-the-art methods, including \mona.
Let us emphasize that, like with \mona, optimizations play a crucial role for
the efficiency of our tool---without them, the basic approach is much less
efficient.
%
Let us briefly mention some of the further possible optimization opportunities.
First, our use of BDDs is not optimal since \mona\ gives us efficient
$\mathit{post}$ only. 
We would benefit from an explicit procedure producing automata encodings with
efficient $\pre$.  
Also, as we mention in Section~\ref{sec:Experiments}, 
performance of our tool could be improved  by a specialised treatment of Boolean variables. 
A plausible solution is to integrate our approach with SAT/SMT technology or
to adapt techniques of \mona. 
%
%
A further logical step would be to use abstraction over the language terms. 
%
%
%
%
Handling more complex formulae in the logic M2L(str), such as those mentioned in Sec.~\ref{sec:Experiments}, also requires specific optimizations.  
Last, we wish to generalize the approach for the logic WS2S, which has
many practical applications too.

\smallskip\noindent\emph{Acknowledgement.}
We thank the anonymous reviewers for their helpful comments on how to improve the presentation in this paper.
This work was supported by the Czech
Science Foundation (projects 14-11384S, 16-17538S, and 16-24707Y), the BUT FIT project
FIT-S-17-4014, and the IT4IXS: IT4Innovations Excellence in Science project (LQ1602).


\bibliographystyle{splncs}
\bibliography{bibliography}



\end{document}

%% file: macros.tex
\newcommand{\td}[1]{\textcolor{blue}{\ifmmode \text{[#1]}\else [#1] \fi}}
\newcommand{\tv}[1]{\textcolor{red!60}{\ifmmode \text{[TV: #1]}\else [TV: #1] \fi}}
\newcommand{\tsf}[1]{\textcolor{green!90}{\ifmmode \text{[TsF: #1]}\else [TsF: #1] \fi}} 
\newcommand{\hide}[1]{}

\newcommand{\aut}[0]{\mathcal{A}}
\newcommand{\autof}[1]{\aut_{#1}}

\newcommand{\pre}[0]{\mathit{pre}}

\newcommand{\lang}{\mathcal{L}}
\newcommand{\langof}[1]{\lang(#1)}
\newcommand{\nat}{\mathbb{N}}

\newcommand{\sing}{\mathrm{Sing}}
\newcommand{\singof}[1]{\sing(#1)}

\newcommand{\projofbase}[1]{\projsymbol_{#1}}
\newcommand{\projof}[2]{\projofbase{#1}\!\left(#2\right)}

\newcommand{\zerosymb}[0]{\bar{0}}

\newcommand{\projsymbol}[0]{\pi}

\newcommand{\cylindr}{\pi}
\newcommand{\cylindrof}[2]{\cylindr_{#1}(#2)}

\newcommand{\powerof}[1]{2^{#1}}

\newcommand{\freeof}[1]{\mathit{free}(#1)}
\newcommand{\cyllangof}[2]{\lang^{#1}(#2)}

\newcommand{\bintrack}[4]{{\resizebox{!}{3.0mm}{$\begin{array}{rl} #1:&#3\\ #2:&#4\end{array}$}}}
\newcommand{\bintrackbr}[4]{{\resizebox{!}{3.0mm}{$\left[\hspace{-0.6mm}\begin{array}{rl}{#1}:&{#3}\\{#2}:&{#4}\end{array}\hspace{-0.6mm}\right]$}}}

\newcommand{\bintracknolabbr}[2]{{\resizebox{!}{3.0mm}{\ensuremath{\left[\hspace{-0.6mm}\begin{array}{l} {#1}\\{#2}\end{array}\hspace{-0.6mm}\right]}}}}
\newcommand{\finst}[1]{F_{#1}}

\newcommand{\preof}[2]{\preofred{#1}(#2)}

\newcommand{\preofred}[1]{\pre{\scriptstyle[#1]}}

\newcommand{\term}{t}

\newcommand{\termof}[1]{\term_{#1}}

\newcommand{\langsubseteq}[0]{\sqsubseteq}
\newcommand{\subsumed}[0]{\langsubseteq}
\newcommand{\langsupseteq}[0]{\sqsupseteq}

\newcommand{\termsubsumed}[0]{\langsubseteq_s}
\newcommand{\termsubsumes}[0]{\langsupseteq_s}

\newcommand{\mona}[0]{\textsc{Mona}}
\newcommand{\osel}[0]{\textsc{jMosel}}
\newcommand{\gaston}[0]{\textsc{Gaston}}

\newcommand{\dwina}[0]{\textsc{dWiNA}}
\newcommand{\toss}[0]{\textsc{Toss}}
\newcommand{\traytel}[0]{\textsc{Coalg}}
\newcommand{\veanes}[0]{\textsc{SFA}}

\newcommand{\rar}[0]{\rightarrow}

\newcommand{\secretsbench}{\texttt{Secrets}}
\newcommand{\strandbenchone}{\texttt{Strand2}}
\newcommand{\strandbenchtwo}{\texttt{Strand}}
\newcommand{\uabebench}{\texttt{UABE}}

\newcommand{\oom}[0]{\texttt{oom}}

\newcommand{\timeout}[0]{{\tiny$>_{\text{2m}}$}}

\newcommand{\dwinabenchone}{\texttt{HornTrans}}
\newcommand{\veanesbench}[1]{\texttt{HornLeq} (+#1)}
\newcommand{\veanesnoaltsbench}{\texttt{HornLeq}}
\newcommand{\tossbench}{\texttt{HornIn}}

\newcommand{\dwinabenchfour}{\texttt{SetSingle}}
\newcommand{\uabeparam}{\texttt{Ex8}}

\newcommand{\uabedparam}[1]{\texttt{Ex11(#1)}}

\newcommand{\unsupportedbench}{\texttt{N/A}}
\newcommand{\timeoutbut}[2]{\timeout ($#1$)}
\newcommand{\oombut}[2]{\oom ($#1$)}

\newcommand{\clearwinner}[1]{\textbf{#1}}
\newcommand{\cw}[1]{\textbf{#1}}

\newcommand{\mydot}[0]{\tikz{\filldraw (0,0) circle [radius=0.4mm];}}

\newcommand{\tcup}{\mathop{\vbox{\offinterlineskip\ialign{%
  \hfil##\hfil\cr
  $\mydot$\cr
  \noalign{\kern-3.5pt}
  $\cup$\cr
  }}}}
\newcommand{\tcap}{\mathop{\vbox{\offinterlineskip\ialign{%
  \hfil##\hfil\cr
  $\mydot$\cr
  \noalign{\kern-4.3pt}
  $\cap$\cr
  }}}}
\newcommand{\tminus}{\mathop{\vbox{\offinterlineskip\ialign{%
  \hfil##\hfil\cr
  \mydot\cr
  \noalign{\kern-1.8pt}
  $-$\cr
  }}}}
\newcommand{\tprojof}[2]{\pi_{#1}(#2)}
\newcommand{\underapp}{\succcurlyeq}
\newcommand{\tcmpl}[1]{\overline {#1}}
\newcommand{\textiff}{\mathrel{\text{iff}}}

\newcommand{\textif}{\mathrel{\text{if}}}

\newcommand{\textor}{\mathrel{\text{or}}}
\newcommand{\textand}{\mathrel{\text{and}}}
\newcommand{\textnot}{{\text{not\ }}}

\newcommand{\memterm}[1]{\epsilon\in{#1}}
\newcommand{\something}{\alpha}

\newcommand{\initof}[1]{I(#1)}
\newcommand{\finof}[1]{F(#1)}

\newcommand{\termequiv}{\rar}
\newcommand{\mezera}{\hspace{2.5mm}}

\newcommand{\stepof}[1]{\tikz[baseline,anchor=base]{\node[draw,circle,inner sep=0.2mm]{#1};}}

\newcommand{\allvars}{\mathbb{V}}
\newcommand{\varset}{\mathcal{X}}
\newcommand{\varsety}{\mathcal{Y}}
\newcommand{\termsetminus}[0]{\ominus}

%% file: uabe-pclengal-dig.tex
\tt  a-a   &  \cw{1.71}  &  30\,253  &  \timeout   &  \timeout \\
\tt  ex10  &  \cw{7.71}  & 131\,835  &     12.67   &  82\,236  \\
\tt  ex11  &      4.40   &   2\,393  &  \cw{0.18}  &   4\,156  \\
\tt  ex12  &  \cw{0.13}  &   2\,591  &      6.31   &  68\,159  \\
\tt  ex13  &  \cw{0.04}  &   2\,601  &      1.19   &  16\,883  \\
\tt  ex16  &  \cw{0.04}  &   3\,384  &      0.28   &   3\,960  \\
\tt  ex17  &      3.52   & 165\,173  &  \cw{0.17}  &   3\,952  \\
\tt  ex18  &  \cw{0.27}  &  19\,463  &  \timeout   &  \timeout \\
\tt  ex2   &      0.18   &  26\,565  &  \cw{0.01}  &   1\,841  \\
\tt  ex20  &      1.46   &   1\,077  &  \cw{0.27}  &  12\,266  \\
\tt  ex21  &  \cw{1.68}  &  30\,253  &  \timeout   &  \timeout \\
\tt  ex4   &  \cw{0.08}  &   6\,797  &      0.50   &  22\,442  \\
\tt  ex6   &  \cw{4.05}  &  27\,903  &     22.69   & 132\,848  \\
\tt  ex7   &      0.90   &      857  &  \cw{0.01}  &      594  \\
\tt  ex8   &      7.69   & 106\,555  &  \cw{0.03}  &   1\,624  \\
\tt  ex9   &      7.16   & 586\,447  &      9.41   & 412\,417  \\
\tt  fib   &  \cw{0.10}  &   8\,128  &     24.19   & 126\,688  \\

%% file: strand-new-pclengal-dig.tex
\tt  bs-loop-else            &             0.05  &  14\,469 &  0.04               & 2\,138  \\
\tt  bs-loop-if-else         &             0.19  &  61\,883 &  \clearwinner{0.08} & 3\,207  \\
\tt  bs-loop-if-if           &             0.38  & 127\,552 &  \clearwinner{0.18} & 5\,428  \\
\tt  sl-insert-after-loop    &\clearwinner{0.01} &   2\,634 &  0.36               & 5\,066  \\
\tt  sl-insert-before-head   &             0.01  &      678 &  0.01               &    541  \\
\tt  sl-insert-before-loop   &             0.01  &   1\,448 &  0.01               &    656  \\
\tt  sl-insert-in-loop       &             0.02  &   5\,945 &  0.01               & 1\,079  \\
\tt  sl-reverse-after-loop   &             0.01  &   1\,941 &  0.01               &    579  \\
\tt  sl-search-in-loop       &             0.08  &  23\,349 &  0.03               & 3\,247  \\

%% file: generated-dig.tex
\veanesnoaltsbench  &  \cite{veanes}          &  \oombut{18}{22.89}  &  \clearwinner{0.03}       &  0.08                   &  \timeoutbut{08}{101.48}  &  \clearwinner{0.03}      &  \clearwinner{0.01}      \\
\veanesbench{3}     &  \cite{veanes}          &  \oombut{18}{20.55}  &  \timeoutbut{11}{29.53}   &  0.16                   &  \timeoutbut{07}{47.65}   &  \timeoutbut{11}{39.23}  &  \clearwinner{0.01}      \\
\veanesbench{4}     &  \cite{veanes}          &  \oombut{18}{20.52}  &  \timeoutbut{13}{29.98}   &  \clearwinner{0.04}     &  \timeoutbut{06}{51.20}   &  \timeoutbut{11}{64.34}  &  \clearwinner{0.01}      \\
\tossbench          &  \cite{ganzow:new}      &  \oombut{15}{7.26}   &  \timeoutbut{11}{54.74}   &  0.07                   &  \timeoutbut{08}{68.05}   &  \timeoutbut{08}{32.44}  &  \clearwinner{0.01}      \\
\dwinabenchone      &  \cite{fiedor:tacas15}  &  86.43               &  \timeoutbut{14}{88.90}   &  \unsupportedbench      &  \unsupportedbench        &  38.56                   &  \clearwinner{1.06}      \\
\dwinabenchfour     &  \cite{fiedor:tacas15}  &  \oombut{04}{1.49}   &  \timeoutbut{08}{38.99}   &  0.10                   &  \unsupportedbench        &  \timeoutbut{03}{40.27}  &  \clearwinner{0.01}      \\
\uabeparam          &  \cite{uabe}            &  \oombut{08}{7.68}   &  \unsupportedbench        &  \unsupportedbench      &  \unsupportedbench        &  \unsupportedbench       &  \clearwinner{0.15}      \\
\uabedparam{10}     &  \cite{uabe}            &  \oombut{14}{15.93}  &  \unsupportedbench        &  \unsupportedbench      &  \unsupportedbench        &  \unsupportedbench       &  \clearwinner{1.62}      \\